\documentclass{elsarticle}

\usepackage{lineno}
\modulolinenumbers[5]

\usepackage{amsmath,amssymb,amsthm}
\usepackage{graphicx}
\usepackage{verbatim}
\usepackage{tabularx}
\usepackage{hyperref}
\usepackage{natbib}
	\setcitestyle{square,comma,numbers,sort&compress}
\usepackage{caption}
\usepackage{subcaption}
\usepackage[margin=1in]{geometry}
\usepackage{mathtools}
\usepackage[none]{hyphenat}
\newcolumntype{L}[1]{>{\raggedright\arraybackslash}p{#1}}
\journal{}

\begin{document}

\begin{frontmatter}

\title{\textbf{Opportunistic Edge Computing:\\ Concepts, Opportunities and Research Challenges}}



\author[mymainaddress]{Richard Olaniyan}
\ead{richard.olaniyan@mail.mcgill.ca}
\author[mymainaddress]{Olamilekan Fadahunsi}
\ead{olamilekan.fadahunsi@mail.mcgill.ca}
\author[mymainaddress]{Muthucumaru Maheswaran\corref{mycorrespondingauthor}}
\cortext[mycorrespondingauthor]{Corresponding author}
\ead{maheswar@cs.mcgill.ca}
\author[mysecondaryaddress]{Mohamed Faten Zhani}
\ead{mfzhani@etsmtl.ca}

\address[mymainaddress]{School of Computer Science\\
Department of Electrical and Computer Engineering\\
McGill University\\
3480 University Street\\
Montreal, Quebec H3A 2A7, Canada}

\address[mysecondaryaddress]{Department of Software and IT Engineering\\
\'Ecole de Technologie Sup\'erieure (\'ETS)\\
1100 Notre-Dame Street West\\
Montreal, Quebec H3C 1K3, Canada}


\begin{abstract}

The growing need for low-latency access to computing resources has motivated the
introduction of edge computing, where resources are strategically placed at the
access networks. Unfortunately, edge computing infrastructures like fogs and
cloudlets have limited scalability and may be prohibitively expensive to install
given the vast edge of the Internet. In this paper, we present Opportunistic
Edge Computing (OEC), a~new computing paradigm that provides a framework to
create scalable infrastructures at the edge using end-user contributed
resources. One of the goals of OEC is to place resources where there is high
demand for them by incentivizing people to share their resources. This paper
defines the OEC paradigm and~the~involved stakeholders and puts forward a
management framework to build, manage and monitor scalable edge infrastructures.
It also highlights the key differences between the OEC computing models and the
existing computing models and shed the light on early works on the topic. 
The paper also presents preliminary experimental results that highlight the benefits and the limitations of OEC compared to the regular cloud computing deployment and the Fog deployment. It finally summarizes key research directions pertaining to~resource management in OEC environments.

\end{abstract}


\end{frontmatter}

\section{Introduction}

Cloud computing in its current form is powered by strategically located
datacenters~\cite{Coady2015DistributedCC} with finite but very large aggregate
resource capacities (CPU, processing power, RAM, storage). For example, Amazon
EC2 which has a 31\% share of the cloud computing market as of 2016 has 14 sites
across the Internet. Emergence of various forms of ``tactile'' Internet
applications such as augmented reality or virtual reality (e.g., Oculus Rift)
and bandwidth hungry applications (e.g., 4K streaming or 4K gaming) is making it
important to place servers closer to the edge so that user applications can
receive responsive performance. This has motivated the need for rethinking cloud
computing architectures with a focus on distributing the infrastructure towards
the edge of the Internet~\cite{satyanarayanan2009case,elkhatib2016using}.

Since its proposal in 2012, Fog computing~\cite{Bonomi:2012:FCR:2342509.2342513}
has been a leading notion of edge computing. In~\cite{chiang}, Chiang postulates
four reasons for Fog computing: (i) real-time processing and cyber-physical
system control, (ii) awareness of client-centric objectives, (iii) pooling of
local resources, and (iv) rapid innovation and affordable scaling. With Internet
of Things (IoT), devices need quick resolution of their analytic computation
requests to meet their quality of service (QoS) requirements. Also, devices
could generate large amount of contextual data (e.g., videos of surroundings in
autonomous cars) that needs real-time processing. To meet the performance
constraints of these applications resources need to be placed along the edge and
data transit times have to be minimized. The edge also has a very large number
of different types of resources. By pooling those resources, computing systems
of different capabilities could be created at very low cost and as and when
needed.

This paper presents an architecture for an Opportunistic Edge Computing (OEC)
framework that dynamically creates pools of resources that are located closer to
the end users. The OEC systems are centrally managed much like traditional cloud
computing systems so that it can carefully manage the QoS delivered to the
customers. However, the resource base used by OEC is distributively owned and
leased to the OEC framework on short-term contracts.  For example, OEC could
lease computing resources from connected cars parked in a parking garage to
create an opportunistic micro-data center that can be closer to end users in the
vicinity. The lease price offered by the OEC would be dependent on the
anticipated future demand for resources in the locality. By varying the lease
price, the OEC can adjust the supply levels to meet the customer demands at the
required QoS levels. Following the same argument offered by cloud computing, the
resources leased opportunistically should be cheaper than the resources obtained
from traditional data centers~\cite{kuada2011social}, when both capital and
operational expenditures are taken into consideration.

The contributions of this paper are as follows:
\begin{itemize}
    \item
    We provide first a motivation for Opportunistic Edge Computing (OEC) by highlighting the technical and social trends for a new framework for opportunistically pooling resources at the edge of the Internet.

    \item
    We present a definition of the Opportunistic Edge Computing model and describe its potential benefits, characteristics, associated business model and potential management framework.

		\item We summarize early work on OEC.
		It is worth noting that existing surveys do not focus on the opportunistic aspect of building edge computing clouds. For instance, surveys by Ahmed et~Al.~\cite{AhmedISCO2016, AhmedCOMMAG2017}, ~Abbas et~Al.~\cite{AbbasMECSurvey2018}, Mach et~Al.~\cite{Mach2017} Guan et~Al.~\cite{GuanICIS2011} provide useful use-cases and applications scenarios for mobile edge computing and identify relevant challenges pertaining to application offloading to the edge.
		Mao et~Al. \cite{MaoIEEESurvey2017} cover existing literature on joint radio-and-computational resource management in mobile edge computing and discuss potential system deployment models as well as cache, mobility, and privacy management issues. Roman et~Al. survey~\cite{ROMANFGCS2018} focuses on security and privacy concerns related to fog computing.
		Unlike these surveys, this work will focus on architectures and frameworks aiming to build opportunistic clouds at the edge of the network using available edge infrastructures and devices.

    \item We provide a taxonomy to classify existing opportunistic computing systems and uses it to
    compare OEC with existing edge-based systems.
 
		\item We conduct several experiments to highlight and discuss the benefits and limitations of Opportunistic Edge Computing compared to the regular cloud computing deployment and the Fog deployment.

    \item We identify the key research challenges that still need to be addressed to implement OEC.
\end{itemize}

The remainder of this paper is structured as follows.
Section~\ref{motivation} motivates the need for a new framework for opportunistically
pooling resources at the edge of the Internet. Section~\ref{OEC}
explains OEC in depth by including a definition, architectural sketch, and business model.
A taxonomy for describing opportunistic computing systems is presented in Section~\ref{classification}.
The OEC is contrasted with existing edge computing systems in Section~\ref{edgecomp}.
In Section~\ref{experiments}, we compare the performance of
an OEC configuration with cloud computing and dedicated fogs using a container-based
emulator for edge computing.
Section~\ref{challenges} provides the research challenges that should be
addressed to realize OEC.

\section{Motivation}
\label{motivation}

\begin{figure}[ht]
\centering
\includegraphics[width=0.8\textwidth]{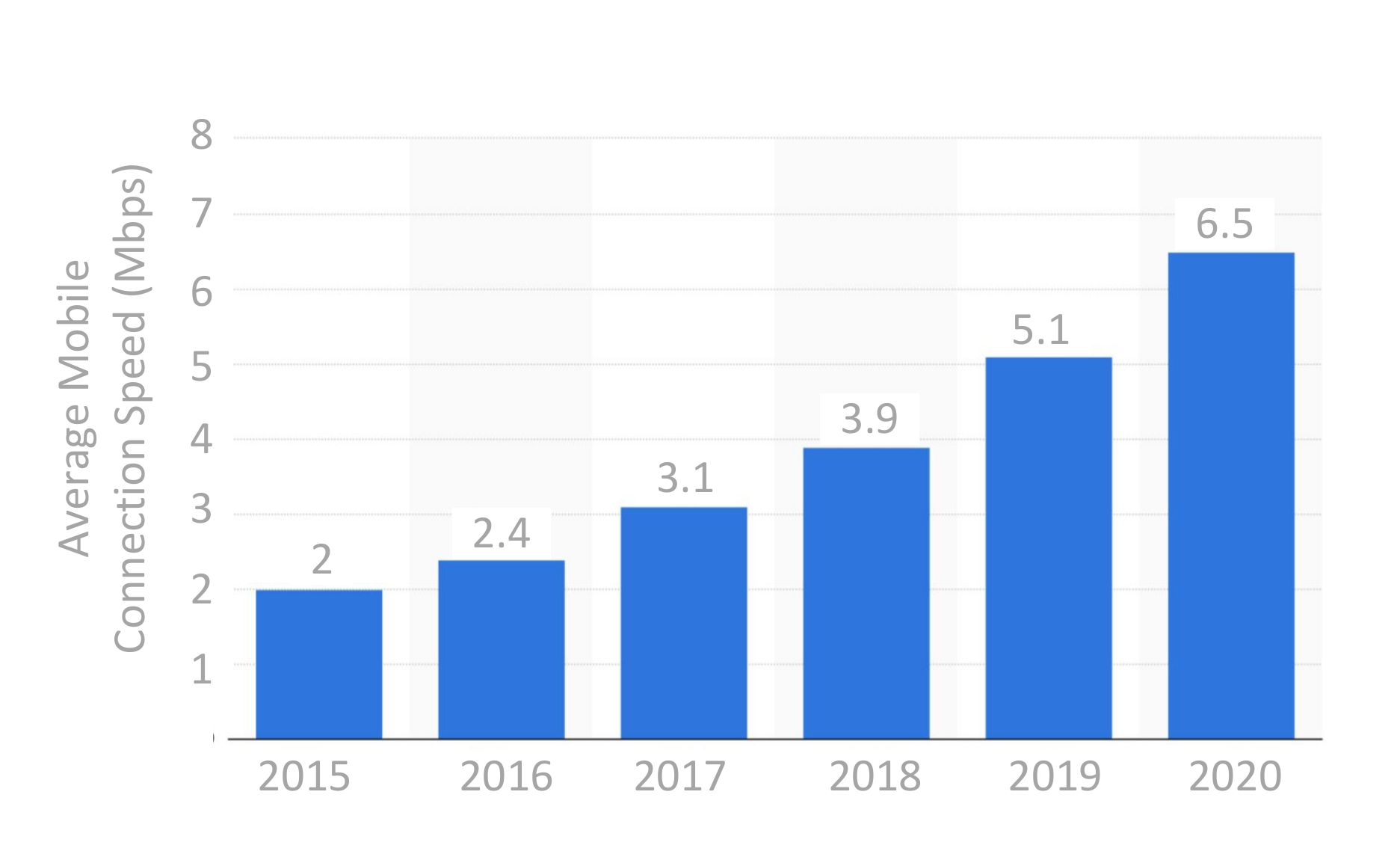}
\caption{Projected improvement in mobile bandwidth (source: Statista)}
\label{fig:mobilespeed}
\end{figure}

There are five major factors that motivate the need for a new paradigm. Firstly,
the {\em cloud is distant}. Long latencies over Wide-Area Networks (WANs) are
already recognized as an impediment for performing
resource- and data-intensive computations with cloud computing
systems~\cite{satyanarayanan2009case}. So, we
need mechanisms to transparently ameliorate the impact of long latencies on
cloud computations. The bandwidth of the mobile access link, which is becoming
the dominant access network in the Internet, is predicted to triple between now
and 2020 as shown in Figure~\ref{fig:mobilespeed}. With high latencies, the
completion times of interactive applications will be dominated by the latencies
and end users would not benefit from the availability of abundant bandwidth.

Secondly, placing serving resources at the edge of the Internet,
which is the core idea
behind several solutions like Cloudlets
~\cite{satyanarayanan2009case,asrani2013mobile,elkhatib2016using} is quite challenging.
In particular, the server resources are placed at or near access routers and are setup with software stacks to quickly launch virtual machines to serve the applications.
A static placement of resources is not suitable for covering dynamically changing
demand hotspots along the edge of the Internet. Also, the approach is not scalable
to handle varying levels of demands for the services.

Thirdly, with the introduction of resource-intensive popular applications such
as 4K gaming and virtual reality, high-performance personal computers will be widely deployed
at end-user premises. High performance and symmetric networks such as
fiber-to-the-home or increasing mobile bandwidth (see Figure~\ref{fig:mobilespeed})
will make such high-performance resources a potent distributed compute cluster.
Without a proper framework for pooling and orchestration of
available resources lot of potential compute power can go to waste.

Fourthly, energy consumption of datacenters is a daunting problem that has
gained significant attention from the research community. Instead of scaling the
capacity of the datacenters to meet increasing demand, peering among datacenters
or using opportunistic resources to augment the resource capacity of the
datacenters could be a way to scale up the datacenter capacity to meet transient
overload conditions; particularly, with edge computing benefits.

\begin{figure}[ht]
\centering
\includegraphics[width=0.99\textwidth]{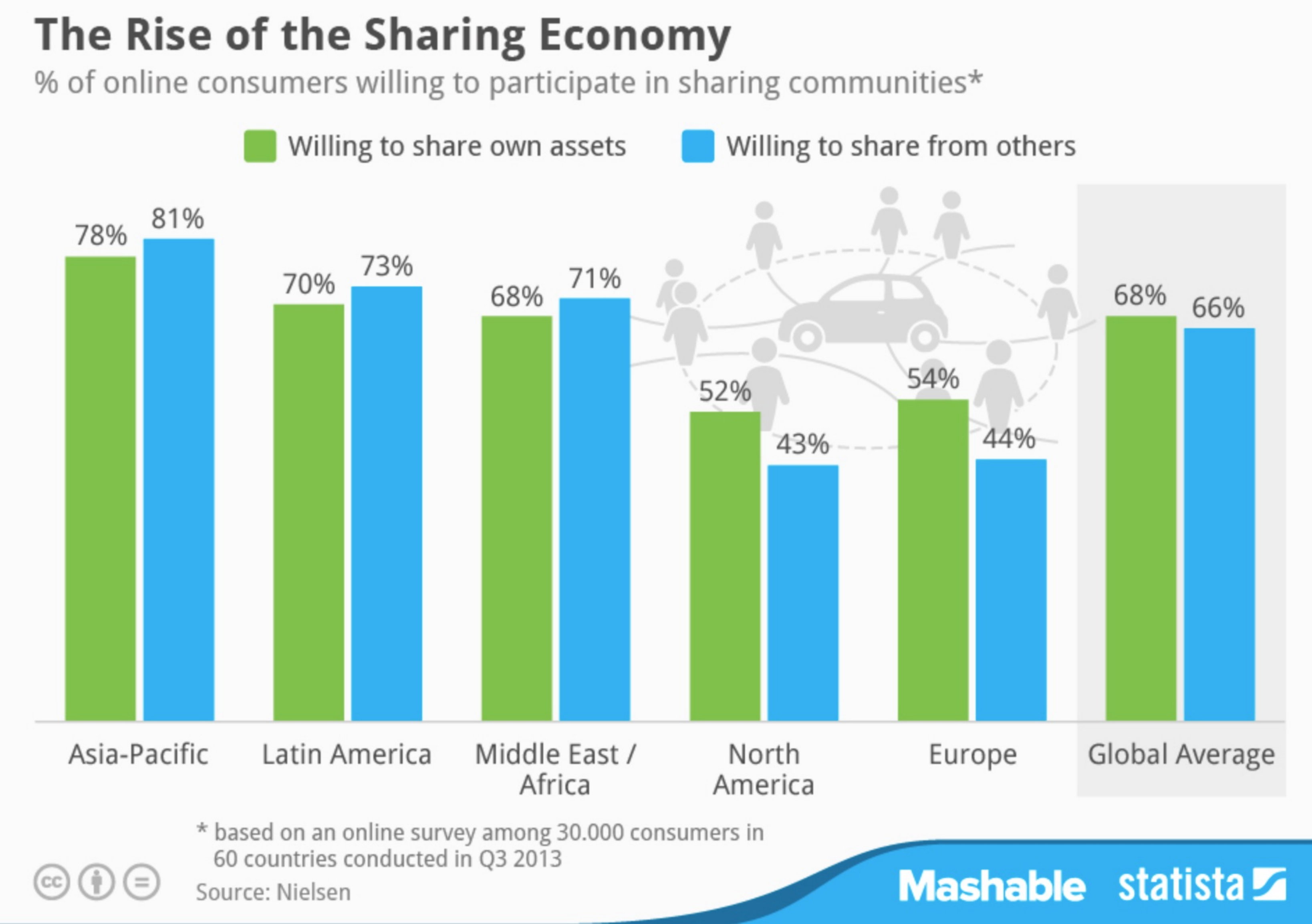}
\caption{User attitude towards resource sharing (source: Statista)}
\label{fig:sharing}
\end{figure}

Lastly, in recent years, {\em sharing} has emerged as
a thriving enterprise on the Internet fueled by a positive user attitude as
shown in Figure~\ref{fig:sharing}.
Popular Internet-based service enterprises have been created to share rooms (e.g., AirBnB)
and rides (e.g., Uber). These services use crowdsourcing to dynamically assemble their resource pools,
which is unlike cloud computing that uses a dedicated pool of resources for resource provisioning.

With resource sharing enterprises that rely on crowdsourcing, incentives are important to keep
the users committed to the system.
One of the ideas used by Uber is a dynamic
pricing model that can increase the price in a locality to balance the supply of
cars with the demands from the passengers. In a ride-sharing system the pricing
model could be changed in a reactive manner but in a high throughput cloud
computing system (particularly with many small tasks), a proactive price
management model is necessary to avoid a pileup of tasks.
AirBnB, on the other hand, creates a marketplace,
where the seller sets the prices for the rooms rented through the service.

\section{Opportunistic Edge Computing}
\label{OEC}

\subsection{Overview}

\begin{figure}[ht]
\centering
\includegraphics[width=0.99\textwidth]{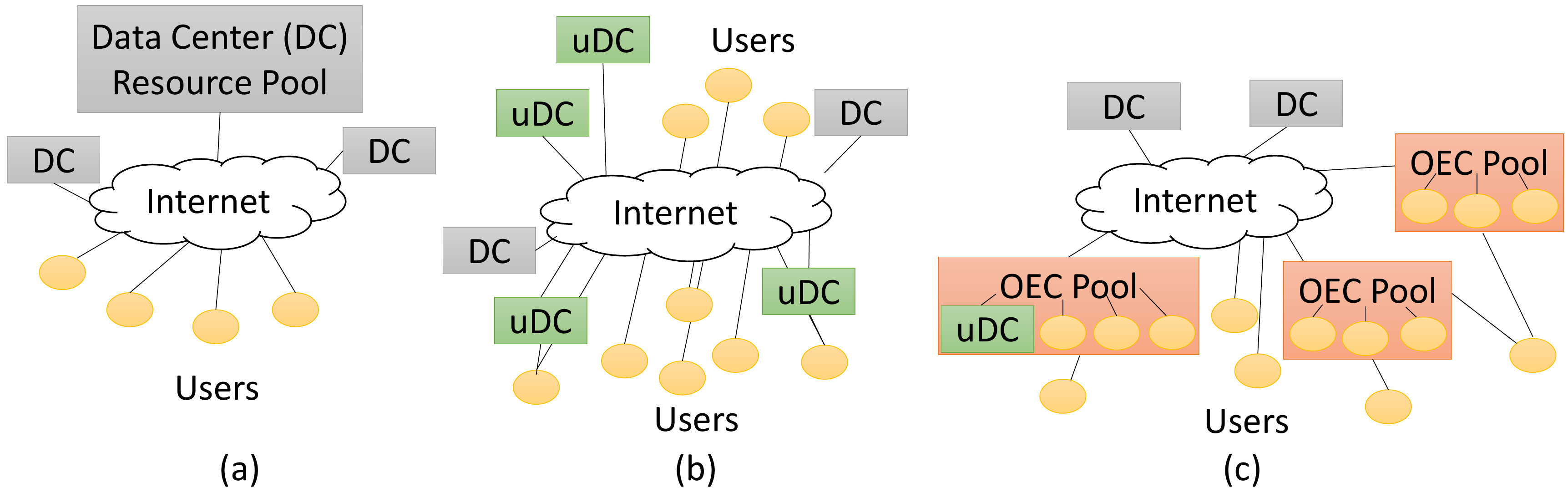}
\caption{Models of cloud computing: (a) core-centric (traditional) clouds,
(b) edge-centric clouds, and (c) opportunistic edge clouds}
\label{fig:sysmodel}
\end{figure}

The best way to understand the Opportunistic Edge Computing (OEC)
model is by contrasting it with existing or
previously proposed cloud computing models. In Figure~\ref{fig:sysmodel}, we
show the system models for three types of clouds: (a) core-centric (traditional), (b)
edge-centric (hyper-distributed) and (c) opportunistic.
The traditional clouds are powered by few
strategically located datacenters.
A global-scale cloud provider (e.g., Amazon) has several datacenters
for redundancy, data localization and geographical loadbalancing.
In this model, the datacenters are placed close to the core of the Internet so that demands
from any point on the Internet can be covered by several datacenters.
The problem
with this model is the {\em distant} compute resources. Any end user application that
relies on cloud computing systems of this type needs to cross the Internet. With
the variable latencies associated with crossing the Internet, this model is {\em not}
considered suitable for time-critical control of IoT or delivering stringent
qualities of services to mobile applications~\cite{satyanarayanan2009case}.

One way of addressing the problems associated with the core-centric
(traditional) clouds is to use ``hyper distribution'' that places micro datacenters at many locations
throughout the Internet~\cite{Coady2015DistributedCC}. In such an edge-centric cloud
configuration, an end user can be connected to a nearby micro datacenter for maximum performance.
While edge-centric cloud computing brings several benefits such as
increased redundancy, better data localization, and much smaller and stable latencies, it
also poses major challenges. One of them is the exponentially increasing management overhead
with the number of sites. Therefore, to achieve the benefits of edge-centric clouds,
all management and deployment challenges need to be solved so that this
model can be competitive with the core-centric cloud model. Further, the cloud service providers
need to be convinced that an edge-centric cloud approach makes a better business case.

The opportunistic edge clouds shown in Figure~\ref{fig:sysmodel}(c) uses a different
approach to push computing to the edge. Instead of locating small resource pools
at strategically selected edge locations (like Cloudlets and Fogs), OEC uses a
broker to create pools of resources at the edge using end-user contributed
resources. So by design the OEC pools are at the edge and can be used to bring
cloud computing to the edge. Because the resources in the OEC pool are end-user
contributed, they are not permanent. One way of addressing their impermanent
nature is for the OEC broker to lease them for a predefined duration of time.
The OEC broker needs to manage the resources in the OEC pool taking the lease
durations into consideration. The end users who lease the resources are
committed to keeping it with the OEC for the agreed upon lease duration. Any
resource that is withdrawn by the end user before the lease duration is
considered as a faulty resource. Also, application hosting on OEC resources need
to follow a soft-state approach like cloudlets and fogs so that OEC resource failures
do not lead to irrecoverable loss of data.

\subsection{Definition, Benefits and Characteristics}

\noindent{\bf Definition:} Opportunistic Edge Computing (OEC) is a model for
providing an on-demand instant access to a shared pool of computing resources
(e.g., servers, storage, networking, applications and services)
opportunistically built using nearby infrastructures (e.g., locally available
equipment and devices). Different from traditional cloud infrastructures, an OEC
infrastructure is discovered, built and managed on the fly depending on the
current availability of devices and infrastructures willing to share their
resources.

Following are some of the key benefits and characteristics of OEC as compared to
existing cloud computing or proposed edge computing paradigms.

\begin{itemize}

\item Proximity to end-users: proximity to end-users is a major advantage
offered by the OEC model. Deployed applications and services will benefit from a
near-zero access latency to end-users and can be instantly and dynamically
provisioned and deployed to serve nearby end-users.

\item Ad hoc infrastructure: OEC allows to create a cloud infrastructure on the
fly with minimal pre-existing infrastructure. As end-users could be also
participants at the same time, they can provide the resources themselves to the
services and applications they are using.  An interesting use-case would be the
attendants of a social or sport event who are dedicating some of their devices'
computing power to services and applications offered during the event.
In this case, users themselves are fueling their services with their own resources.

\item Auto-scalability: another key benefit of OEC is that the amount of pooled
resources is proportional to the number of participants. Intuitively, the demand
for a service scales with the number of end-users. If these users are making their
 resources available to the OEC, the capacity of the opportunistic edge cloud will
 naturally scale up and down depending on the number of users and the demand.

\item Multi-tenancy: the resources of the opportunistic edge cloud can be leveraged
to accommodate the requirements of multiple service providers. In this case, a resource
management module should defines the way resources are sliced and distributed among the SPs
as well as the level of resource isolation depending on the performance objectives and requirements.

\item Improved cost-effectiveness: a large adoption of the OEC and a high number of
participants will lead to a better economy of scale, incurring lower deployment
costs for services and applications. At the same time, end-users
might be themselves participants of the opportunistic edge cloud and, hence
they can obtain the benefit of the OEC resources and services at lower prices.
Finally, OEC reduces the need for potentially expensive resources offered by
traditional cloud providers.

\end{itemize}

\subsection{Business Models}

\begin{figure}[htbp]
	\centering
		\includegraphics[width=0.60\textwidth]{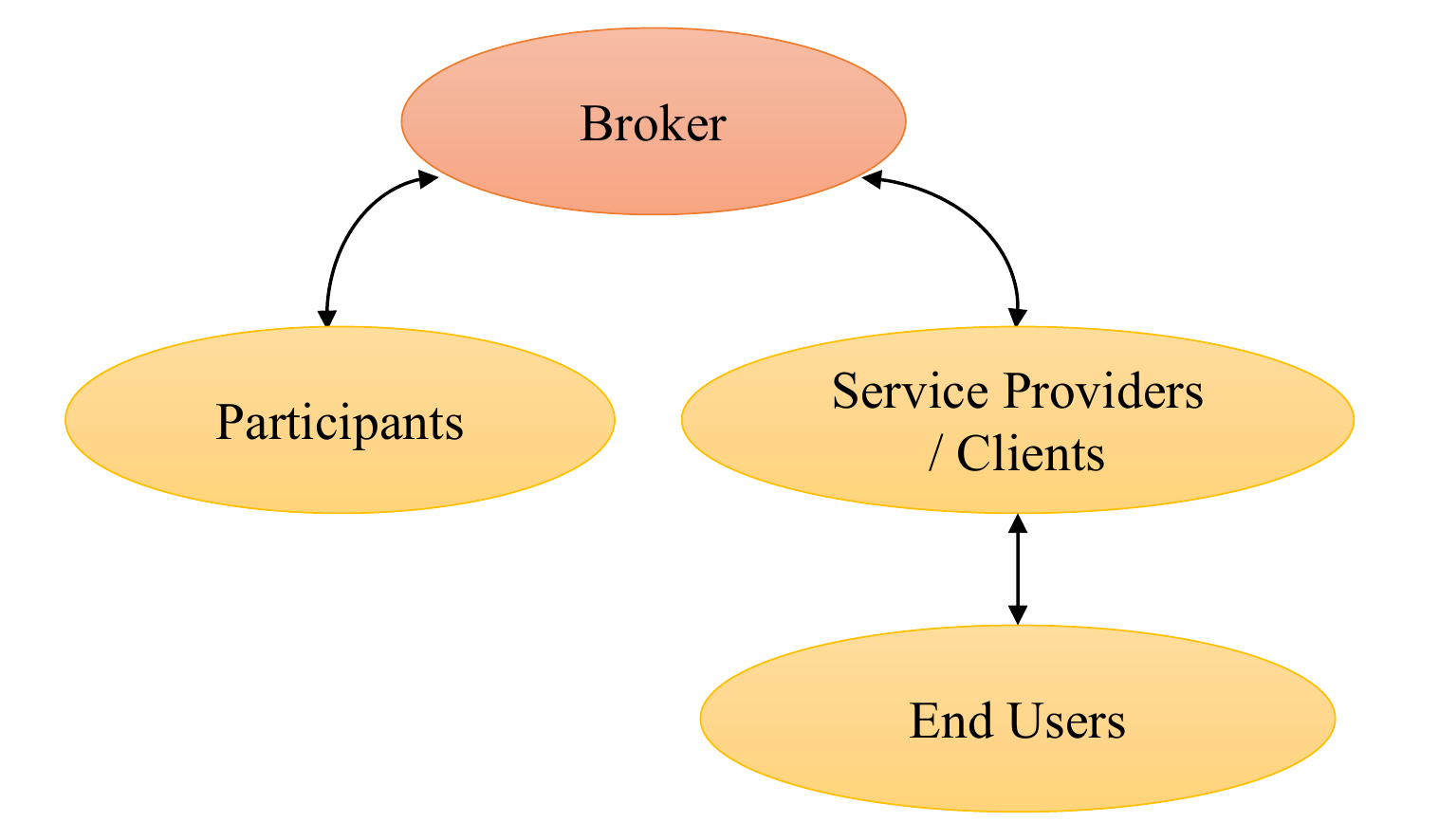}
	\caption{OEC business model}
	\label{fig:BusinessModel}
\end{figure}

We believe the OEC business model should involve the following four
stakeholders as shown in Figure~\ref{fig:BusinessModel}:

\begin{itemize}

\item Participant: participants own the physical resources (e.g., LANs, desktop
computers, mobiles, laptops).  Motivated by some incentives, they put these
resources under the control of a broker.

\item Broker: brokers are in charge of building and managing the opportunistic edge cloud
using resources offered by the participants. A broker can be a company that negotiates
resource acquirement contracts with participants, pools their
resources, or slices it and offers it to Service providers (SPs). The broker can
also use these resources to run jobs as requested by potential clients.

\item Service Provider/Client: SPs or clients lease resources from a broker and
use it to deploy their applications or services.
The clients may choose not to lease resources but rather request the broker to
run jobs on the opportunistic edge cloud

\item End-user: end-users are the users of applications and services offered by~SPs.

\end{itemize}

\subsection{OEC Framework}\label{OECFramework}

\begin{figure}[ht]
\centering
\includegraphics[width=0.80\textwidth]{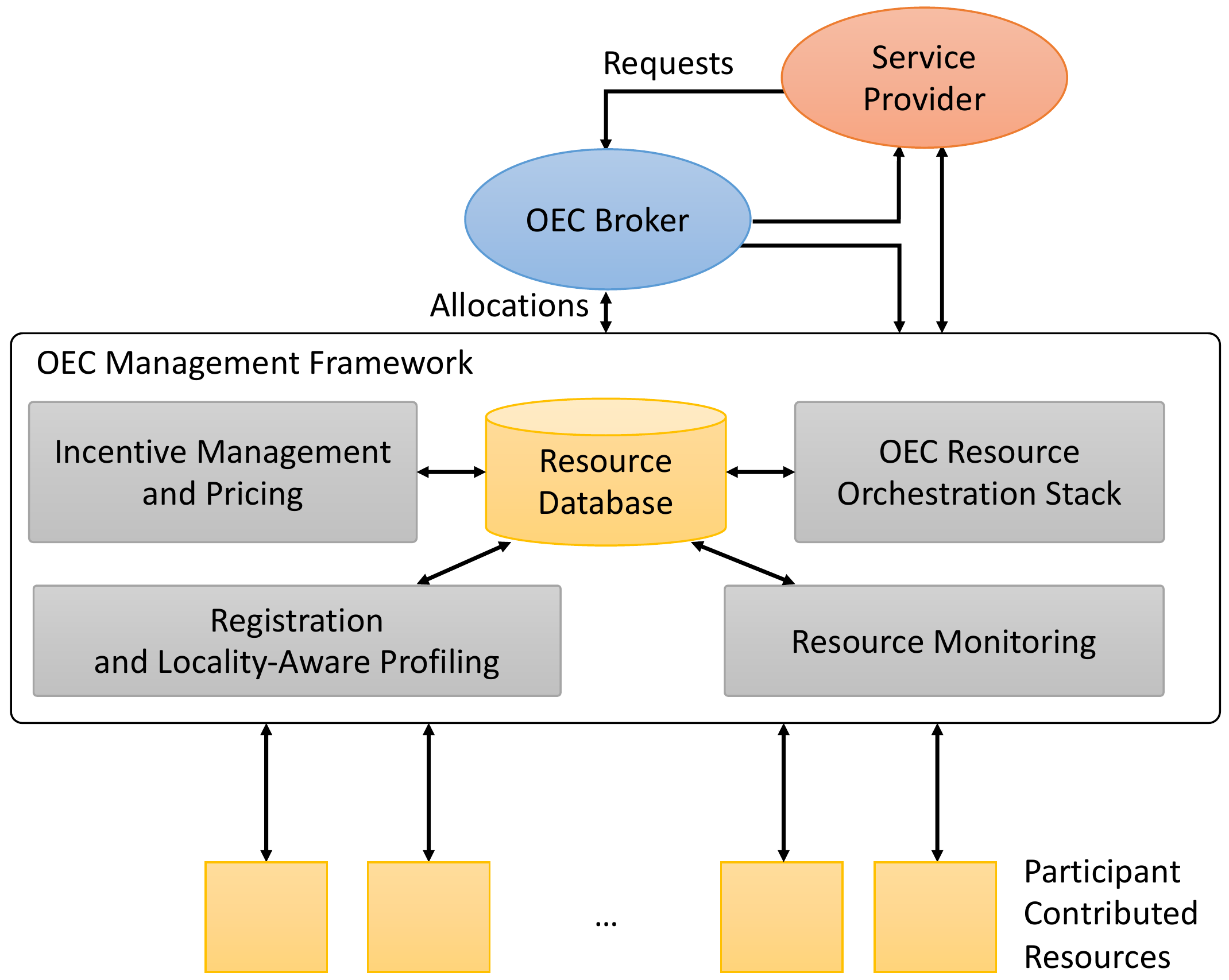}
\caption{Key architectural elements of the OEC framework}
\label{fig:arch}
\end{figure}

The key architectural elements of the OEC framework are shown in Figure~\ref{fig:arch}.
The participants contributed resources are registered with the system using the registration
and the Locality-Aware Resource Profiler. For each resource, this module is responsible for
determining the resource type and characteristics (e.g., amount, performance, availability time,
locality, reliability). Using this information, OEC estimates the demand for a resource in the
given network vicinity and the level of incentive that can be offered to a participant to
donate the resource for a specific time period.

The incentive management and pricing module is responsible for managing and determining the
incentives offered to participants and eventual prices for their contributed resources.
It can use historic data on the supply of resources from the participants to negotiate
the incentives, the resource prices and the engagement plan with the OEC.
For instance, this module could help participants to determine how much they could earn in
incentives when leasing their resources when they are not using them (e.g., home computers when people are at work or school).

OEC should implement mechanisms to support multi-tenancy over the resources provided by the participants.
This is achieved through the OEC Resource Orchestration Stack which assembles the offered
resources and divide it into isolated slices so that  different service providers could use them
to deploy their services and applications or to run computing jobs. This module is also in charge
of managing faults and the intermittence of the participants' devices in order to ensure that the
requested resources is always available.

The resource monitoring module is responsible for continuously monitoring the participant's
resources and to provide relevant statistics for the other modules (e.g., utilization,
availability, failures, performance).
Finally, the resource database keeps track of all these statistics, along with all relevant
information about the participants, the service providers/clients, request description,
and the allocated resources.


\section{Related Computing Models and Early Work on OEC}
\label{edgecomp}

A number of cloud-dependent technologies have sprung up in recent years to bring cloud computing closer to end users including Fog Computing and cloudlets, Mobile Cloud Computing, and Vehicular Cloud Computing. 
In this section, we describe these concepts and technologies and compare them. We then describe how they relate to OEC and summarize the early work on OEC.

\subsection{Fog Computing and Cloudlets}

The Fog Computing advocates moving computing closer to end devices by
operating at the edge of the network~\cite{Bonomi:2012:FCR:2342509.2342513}.
Some characteristics of Fogs are: mobility, wireless access, low latency, location
awareness, large number of end devices, real-time interaction, interoperability
and heterogeneity. These characteristics make Fogs ideal for IoT applications and
services. 

\begin{figure}[ht]
\centering
\includegraphics[width=0.8\textwidth]{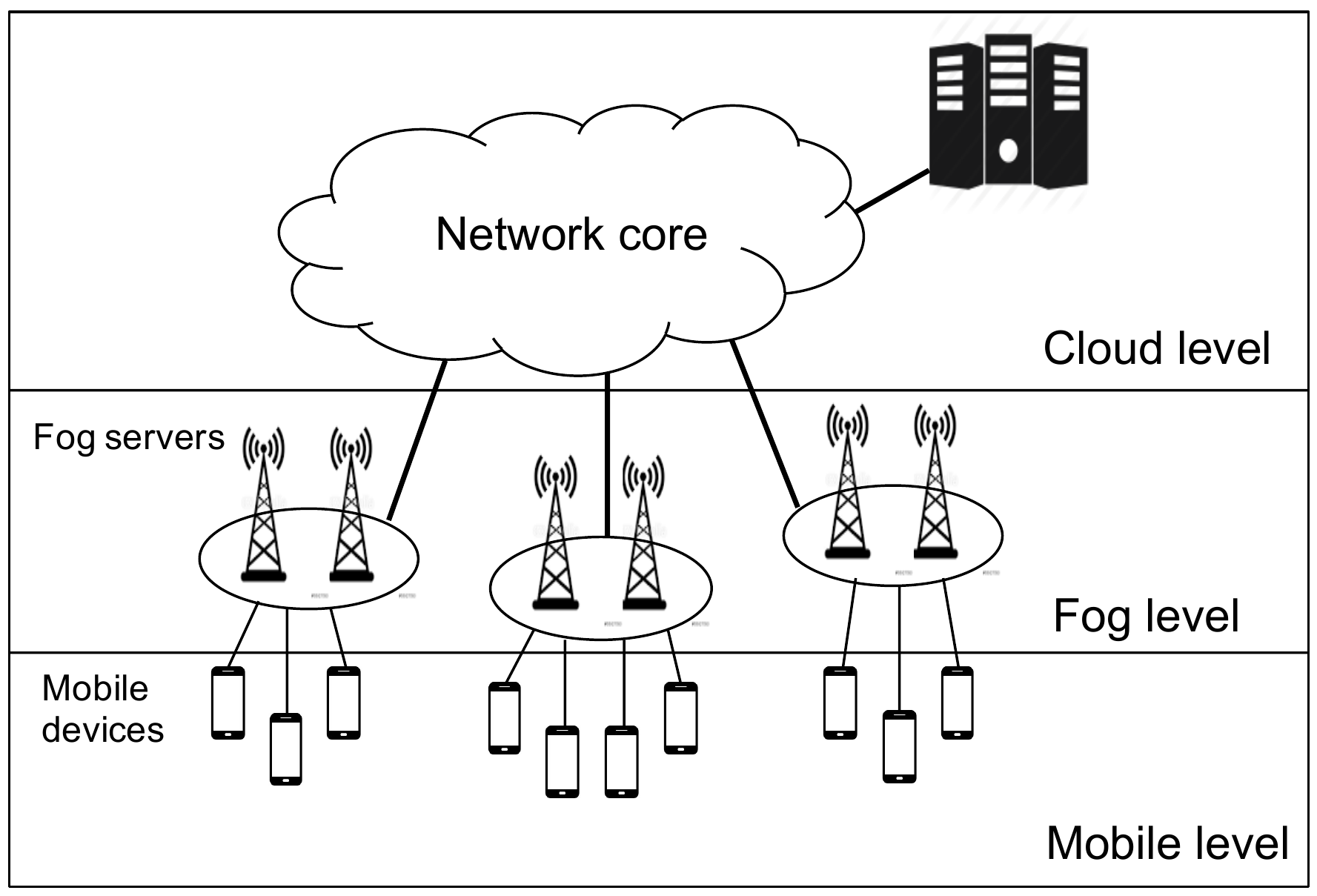}
\caption{General architecture of Fog computing}
\label{fig:fog}
\end{figure}

Some example cases where Fogs have been deployed as a platform for IoT
are Connected Vehicle, Smart Grid, Wireless Sensors and Actuators and Smart
Cities~\cite{Bonomi:2012:FCR:2342509.2342513, SanchezIborra2018199}.
Smart Grid and Smart Cities are also good examples of where the
richness of the Fog and its relationship with the cloud is used to deploy IoT
applications. Fogs operate well with cloud as it can provide local information and resources whereas the cloud 
cloud provides more resources at a larger scale \cite{Bonomi:2012:FCR:2342509.2342513}.
Figure~\ref{fig:fog} shows the general architecture of a Fog. Cloud computing is
extended by the Fog introducing an intermediary layer
between the cloud and mobile devices. Fog servers are deployed in the intermediate layer
in a geo-distributed manner such that they are within the local vicinity of mobile
users~\cite{Luan2016}.

Another close concept is that of Cloudlets. A Cloudlet is a trusted, resource rich computer or cluster of computers that is well connected to the Internet and available for use by nearby mobile devices. It can also be viewed as a small, simple device that resides nearby (e.g., a nearby coffee shop). When needed, the device downloads user data from a centralized location, permitting local access by the user and thereby reducing latency. 
When finished, the user data can be returned to the centralized location, if necessary. This process is transparent to the user, except that the user is
pleased with a faster response \cite{asrani2013mobile}. In \cite{satyanarayanan2009case}, the authors proposed an architecture that allows
to rapidly instantiate customizable services using virtual machines on a nearby Cloudlet and to allow the use of the
services over a wireless LAN.

%

\subsection{Mobile Cloud Computing}

The use of mobile devices have increased drastically over the years, and the mobile technology market has also grown exponentially. Mobile users have increasing volumes of data and information (e.g., pictures, videos) that need to be processed and stored. However, due to storage, processing and battery constraints on mobile
devices, the data need to be stored and processed in a nearby location where it can be easily accessed and modified.  
Mobile Cloud Computing (MCC) was proposed to cater to this need, and it mainly refers to data processing and storage services provided to mobile devices from external infrastructures (e.g., other devices) in order to enhance their processing abilities \cite{Dinh2013,tayade}. 

\subsection{Vehicular Cloud Computing}

The focus of Vehicular Cloud Computing (VCC) is to offer on demand solutions for unpredictable events in a proactive fashion \cite{Hadaller:2007:VOC:1247660.1247685}. VCC relies on a huge fleet of vehicles that criss-cross the roadways, airways and waterways and with most of them having a permanent Internet presence can be thought of as a huge farm of computers on the move. Resources offered by a vehicle can include the computational, storage and sensing resources inherent in the vehicles. Parked vehicles are also a vast unexploited resource which is otherwise wasted. 
The vast number of vehicles on the streets, parking lots, roadways makes them good candidates for nodes in a cloud computing network \cite{Whaiduzzaman:2014:SVC:2608850.2608954}. 
The goal is to leverage on board resources in participating cars. Examples include airport parking spaces where travelers usually park their cars. The parking garage can power the vehicles and make the computing resources available as a parking garage datacenter. The resources in these vehicles can be leveraged to provide third-party or community services. One distinguishing factor between standard nodes in a conventional cloud and the vehicles is \emph{autonomy} and \emph{mobility} attributes that the vehicles possess. For example, drivers stuck in traffic congestion can pool their on-board computing resources to help city traffic planners run complex simulations to remove congestion by rescheduling traffic lights. The VCC has the potential to use  vehicles to cooperatively solve problems that would take a centralized system an inordinate amount of time which thereby renders the solution useless \cite{Abuelela:2010:TVC:1971519.1971522}.

\subsection{Comparison of the Edge Computing Variations}

In this section, we examine the relationships between the different variations
of edge computing based on selected characteristics as shown in Table~\ref{plan}.
\begin{table}[!pth]
\caption{Comparison of cloud computing variations}

\begin{center}
\begin{tabular}{|L{2.5cm}|L{2.5cm}|L{2.5cm}|L{2.5cm}|L{3cm}|}
\hline \textbf{Dimension} & \textbf{Fog Computing / Cloudlets} & \textbf{Mobile Cloud Computing} & \textbf{Vehicular Cloud Computing} & \textbf{Opportunistic Edge Computing}\\
\hline \textbf{Target User} & End-users and mobile users & Mobile users & Cars and mobile users & Service providers\\
\hline \textbf{Resource Provider} & Fog or Cloudlet operator & Cloud service provider & Vehicle owners (mobile users) & End users (participants), Cloudlet operator \\
\hline \textbf{Resource Type} & Storage and compute capacity & All resource offered by cloud & Resources located in the vehicles
& Heterogeneous resources based on participant contribution \\
\hline \textbf{Resource Pool Location} & At the access network & At the cloud & At the edge in vehicles & At the edge in participant computers and 
participating cloudlets \\
\hline \textbf{Resource Capacity Scaling} & Fixed at deployment time & Elastic scaling & Dynamic and depends on vehicle mobility & Self-scaling and depends on user mobility and cloudlet availability \\
\hline

\end{tabular}
\end{center}
\label{plan}
\end{table}
%
%
%
As~can~be~seen~from the table, achieving locality is one of the key objectives for the
different variations of edge computing. The OEC shares this objective; however,
it approaches the problem in a different way.
In vehicular clouds, resources are located with the vehicle.
Similarly, they are located with the user in MCC.
In Fog computing, however, resources are strategically placed at the edge of the
network. For instance, the edge router itself could host large volumes of
resources to cater for the local demand.
So, the access network i.e., the first hop of the network will have
the resources. In Fog computing, the resources capacities are fixed although it
can be argued that capacities are distributed and located at the ``edge of the
network.'' Whereas, VCC and MCC allow the resource pool
to scale and rearrange in composition in a dynamic manner.\par

OEC differs from the emerging cloud computing derivations in the following ways:
\begin{enumerate}
    \item In OEC, the cloud is formed in a dynamic manner and
changes rapidly unlike in Fog Computing where fog servers are deployed at
specific locations.
    \item The procurement and provisioning of resources in OEC are determined by
    the current customer demand for computing resources. Therefore, OEC provides
    an on-demand cloud.
\end{enumerate}

\subsection{Early Work on OEC}


In this section, we cover existing architectures and frameworks proposed in the literature for pooling resources at the edge of the Internet.
%
%
In the following, we divide opportunistic edge computing architectures into two major categories: (1) peer-to-peer architecture where edge devices cooperate among themselves to build and to offload tasks, and (2) centralized architectures or broker-based, where a central entity is in charge of collecting information about edge devices, build an opportunistic cloud and make task scheduling decisions. 


An example of opportunistic peer-to-peer systems is the one proposed by Teo et al.~\cite{TeoHyrax2012} that allows mobile devices that are close to each other to communicate directly and to build a cloud of mobile devices. This enables collaborative data-intensive computing without the need to consume bandwidth towards Internet.
Another example is the one proposed by Mascitti et al.~\cite{mascitti2014service}, where mobile devices directly provide services to each other without necessarily going through a centralized service. Their algorithm takes into account both the mobility of the devices and their computational capabilities to derive stochastic measures useful for comparing the possible alternatives to satisfy a request for service. The algorithm selects the best alternative for providing the service to a client among the available services on~other~mobile~devices. 
Mtibaa et al.,~\cite{Mtibaa2015-TMOC} proposed a peer-to-peer architecture for  opportunistic offloading over heterogeneous devices including mobile, cloudlets, and clouds. In this architecture, each mobile can choose to offload some tasks to other devices based on different objectives like reducing energy consumption and minimizing the task execution times. 
These peer-to-peer architectures have the advantage of building opportunistic infrastructures to leverage nearby resources. However, they assume each mobile is able to collect statistics about nearby devices, build and manage the opportunistic cloud and take offloading decisions. This might incur several limitations impacting the overall performance. Indeed, in addition to the management overhead associated with such tasks (e.g., monitoring, discovering neighboring nodes and building the OEC, scheduling), mobiles are scheduling tasks in an independent manner, which may result in poor performance especially when the same resources are offered concurrently to~multiple~mobile~devices. 
Shi et al.,~\cite{ShiSerendipityMobiHoc2012} proposed Serendipity, a system that enables a mobile devices to communicate with devices and dispatch computational tasks across them with the goal of speeding up task computation time and reducing mobile energy consumption.


The second category is the centralized architecture where a central entity is responsible  for discovering and managing the opportunistic resources.  

Verba et~al.~\cite{VerbaPaaS2017} proposed a messaging-based platform aiming at abstracting communication protocols to allow edge devices to communicate and share their resources. The proposed platform relies on a Gateway Controller to build clusters of edge devices and deploy applications over them.
Goyal et~al.,~\cite{goyal2011collective} proposed to use a collective manager that aggregates volunteer resources and orchestrates their operation to deliver predefined services such as content distribution and backup to the clients. The collective manager has incentive management schemes that operate with complete or incomplete information regarding the resource sharing activities of the end users. 
End users contribute by providing their resources and get paid in a virtual currency for offering their resources in a consistent manner and can use the same currency to use the services offered by the cloud manager. 
Habak et~al.,~\cite{HabakFemtoClouds2015} put forward a system to discover available mobile devices in order to build a femtocloud cluster that can run processing tasks. The proposed system is able to estimate the devices' presence time and computing capabilities and to schedule arriving tasks accordingly.
Farris et~al.,~\cite{FarrisMIFaaS2016,FarrisFederated2015} assume IoT devices can be grouped to form an IoT cloud. Multiple IoT clouds can be then federated to form a larger pool of resources. The authors hence propose a system that operates at the edge in order to manage and federate IoT clouds and allow to run high level services and tasks using the federated resources.



\section{Classification of Opportunistic Computing}
\label{classification}

Opportunistic computing of various types are already happening in the Internet.
In this section, we examine them and classify them based on some important
dimensions. The purpose of this activity is to understand the relationships among
the different forms of opportunistic computing and see how OEC fits in
with the existing opportunistic computing schemes.

One of the dimensions of classification is the level at which opportunistic
computing takes place, which can impact the attainable performance and trust.
Another dimension is the length of the contract (if applicable) that binds the
participants in the computing activity.

\subsection{Opportunistic Computing Levels}


Opportunistic computing at the {\em resource level}
means two nodes are willing to share resources with
each other. For example, allowing another user to spawn a virtual machine in
one's computer based on some incentive exchange is a resource-level
opportunistic computing. Through such opportunistic computing one can run any
computing or data processing activity in the remote machine without being limited by the software
or services offered by the remote machine.


Network level opportunistic computing refers to two nodes willing to provide network layer
data forwarding for each other. Mobile ad hoc networks is an example of such a
network level opportunistic computing, where nodes that are close to each
other are providing  packet forwarding facility to each other. Network level
augmentation has been extended to delay tolerant network as
well~\cite{Keshav:2010:DPR:1836001.1836007,Pal2013,Boldrini2014,Sciancalepore2014,Seth2006},
where physical movements of computing elements are used to create ``store and
carry'' networks. Such networks are still examples of network level
opportunistic computing because the opportunistic actions (i.e.,~packet
forwarding) are limited to the network layer.


In a large-scale computing system, the management level functions play a key role in
dictating the quality of service delivered to the applications. For instance, a
large collection of computers can be managed by a batch processing job scheduler
as a high-performance computing center or as a cloud computing system
using a software stack that virtualizes the resources
and offers slices to customers
according to their needs. The time-to-complete the jobs and the types of QoS
can differ dramatically between the two situations. Similarly, the
composition of the management level functions can significantly impact the QoS
offered by an opportunistic computing system. For instance, resources
pooled from users could be used without any contracts like in peer-to-peer
systems or with short-term contracts. However, with contractual agreements, we
can derive quality-of-service from the same resources.


Most of the working opportunistic computing systems operate at the application
level. Ease of creating opportunistic computing systems at the application level
is one of the primary reasons for this situation. For instance, vast majority of
the crowdsourcing systems operate at the application level. There is a
wide-variety of systems that fall into the crowdsourcing category. Some of them
provide access to data that would otherwise not be available (e.g., peer-to-peer
file sharing or BitTorrent networks) and others provide knowledge or service
(e.g., question answering networks). In addition to crowdsourcing, opportunistic
computing at the application layer has its own share of resource-sharing
systems. The most popular resource-sharing system at the application layer is
the SETI@Home system that allows the pooling of volunteer computing resources
for solving a scientific computing tasks. Although application-layer
opportunistic computing systems are easy to create, they are quite weak in the
quality-of-service guarantees they can provide for their resources or services.

\subsection{A Taxonomy for Opportunistic Computing Systems}

Our classification is based on two dimensions: opportunistic computing level
and the duration of commitment for opportunistic computing.
The opportunistic computing can take place in one of the levels described above.
The commitment can be of two type: with contractual obligations or no obligations.
We use the
length of the contractual obligations as a
parameter here. A zero length contract implies no contract.

\begin{figure}[ht]
\centering
\includegraphics[width=0.9\textwidth]{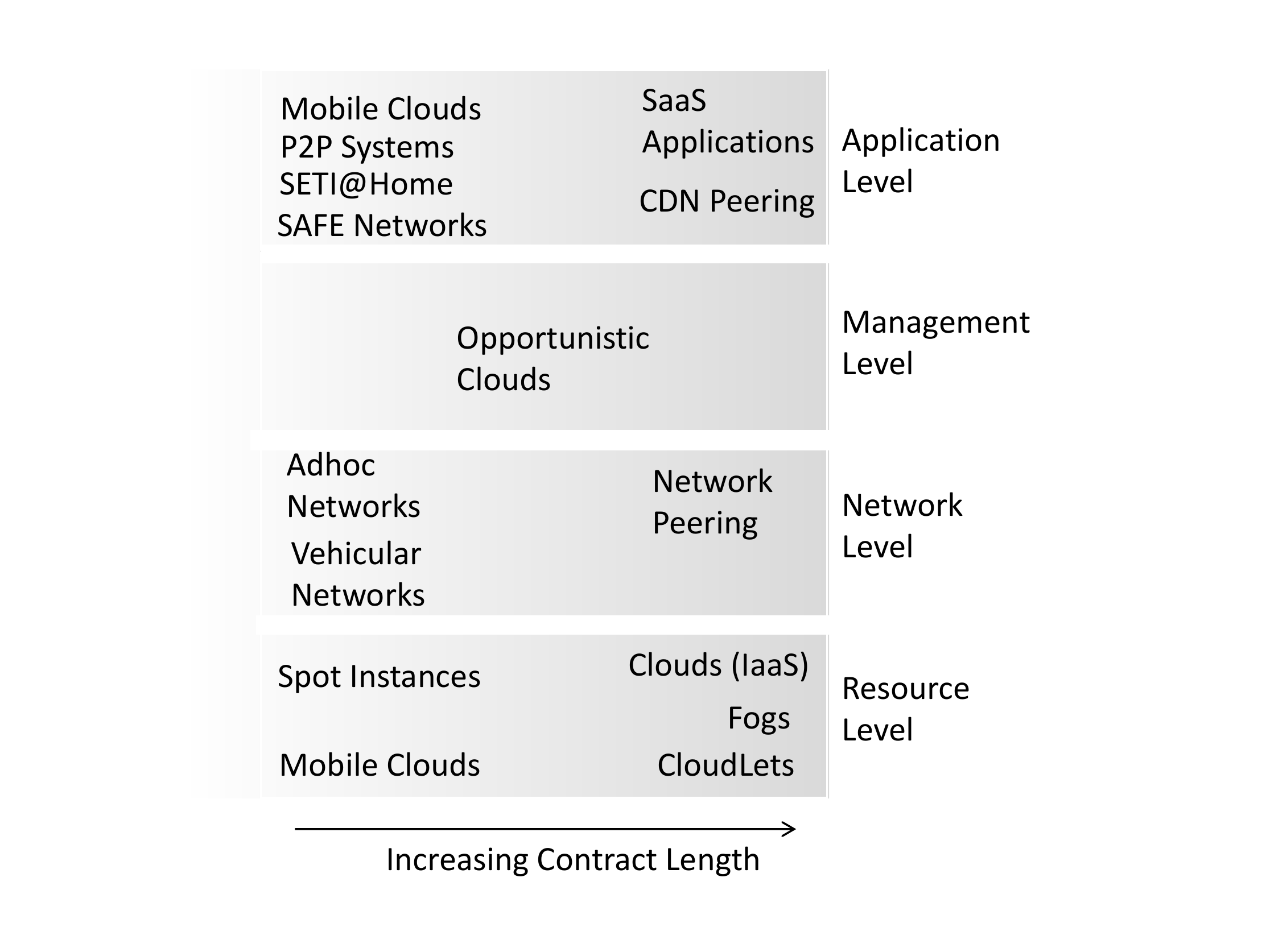}
\caption{A taxonomy for opportunistic computing systems}
\label{fig:tax}
\end{figure}

Figure~\ref{fig:tax} shows the taxonomy and a mapping of known
distributed computing systems (both opportunistic and
non opportunistic) onto the taxonomy.
From Figure~\ref{fig:tax}, we can observe that the application level is the most
popular opportunistic computing layer. Ease of initiating an opportunistic
linkage at the application level is the primary reason for its popularity.  Many
opportunistic computing systems have been developed at the networking level as
well. As expected, these systems strive to provide connectivity over a network
that  is prone to frequent disconnections. End-to-end connectivity creation to
run networking applications is the purpose of this class of opportunistic systems.

Vehicular clouds are another example of opportunistic systems that operate at
the network level. The major objective of vehicular clouds is to provide access
to services and data in a broader locality. Vehicular clouds are
particularly useful for applications
such as traffic control and regulation, car parking, bad road updates and
collision avoidance.

Infrastructure-as-a-Service providers sometimes rent out virtual servers through
auctions at a much cheaper price. These virtual servers are called spot
instances.
In the spot instance market,
reliability and availability are traded off for lower-cost
\cite{Mazzucco2011,Yi2010}. In \cite{Yi2010} a model was proposed to reduce the
effective cost of spot instances via checkpointing, and also in
\cite{Voorsluys2011} a reliable and fault tolerant provisioning and bidding
scheme was proposed for reliable provisioning of spot instances for
compute-intensive applications. Thus, spot instances can be used to get cloud
computing in an opportunistic manner by leveraging the cheap cost at a
particular point in time.


\section{Experimental Results and Discussion}
\label{experiments}

In this section, we run several experiments to highlight the benefits and limitations of Opportunistic Edge Computing compared to the regular cloud computing deployment and the Fog deployment. We hence consider a scenario where several end-user devices have to run computational jobs and they need to offload them. To do so, the devices can resort to one of the following deployments: 
\begin{itemize}
	\item The Cloud deployment, which is the regular cloud infrastructure with limitless resources; However, it is situated far (i.e., non-negligible latency) from the devices. 
	\item The {\em Dedicated Fogs}, which is a collection of fog-level nodes that are permanently installed close to the edge devices. This would be a perfect deployment to provide high performance as it allows a minimal latency with high amount of resources at the edge. However, deploying and maintaining a large number of dedicated fogs at the edge is costly and not possible to implement in practice in a multiple locations. 
	
	\item The {\em OEC Cloud}, which is a collection of nodes (e.g., end-user devices or even fogs) that are opportunistically pooled at the edge. Because dedicated fogs are costlier to place and maintain in the network, the OEC Cloud would be a way of achieving the same performance objectives at a fraction of the cost. However, one of the differences between the dedicated fogs and the OEC Cloud is the disconnections and reconnections that could be be experienced with the OEC Cloud. For example, vehicles plugged into a smart-city electric car charging network could be hosting opportunistic nodes. As the vehicles remain plugged in the fog, nodes could be available for serving requests; However, they become unavailable when the vehicles are unplugged from the charging network.

	
\end{itemize}

\subsection{Experimental Setup}

For the experiments, we developed an edge computing emulator (developed as
part of the JAMScript project\footnote{https://anrl.github.io/JAMScript-beta}). In
this emulator, nodes are represented by Docker containers and networks are represented
by virtual switches. In the current version, the Linux bridge is used to implement virtual
switches. To inject different delays between the different network nodes, the emulator
uses the Pumba Chaos Testing Tool\footnote{https://github.com/alexei-led/pumba}.
Using Pumba, we are able to implement different link delays to emulate wide- and local-area connections.
For instance, devices would be closer to fogs and to other devices in a given locality, but
far away from the cloud and other devices in other localities. 

To implement the idea of locality, we divide the network into zones and distribute the nodes across
the different zones.  We can then identify three types of network latencies: (i) intra-zone latency that
is experienced by devices when they communicate with other devices and fogs in the same zone, 
(ii) inter-zone latency that is experienced when a device communicates
with another device or fog from another zone, (iii) cloud latency to access the cloud.
In our experiments, the intra-zone latency is set to an average of 5ms and 2ms variance, the inter-zone latency is
set to 10ms and 3ms variance, and the cloud latency is set to 50ms and 6ms variance.

Each node is a Docker container. The CPU allocation for the containers are set at the default level and is equal across all containers. 


The devices, fogs, and cloud are organized in a tree in these experiments. That is, the fogs are connected to the cloud and the devices are connected to the
fog. The devices are attached to the different fogs using a distributed selection function. A random matching function is used by the devices to select
their fog for attachment. 

The devices want to offload a linear equation solution problem (i.e., solving $A x = b$) to the fog. The matrix $A$ and vector $b$ are generated by the device
through its sensing operations. Given these values, it wants to find the solution vector $x$ that it can use to perform operations in the local context.
In this problem, both the matrix $A$ and vector $b$ are variables. However, in these experiments, matrix $A$ is uploaded by the device at the beginning of the
experiment and remains unchanged throughout. Once A is uploaded, the end-user devices start to randomly send requests to the used computing platform (i.e., cloud, dedicated fogs or OEC cloud).
Each request has a fresh value of $b$ and seeks a new solution vector $x$.

We use a sparse matrix with 1294 elements (147-by-147) from the University of Florida Sparse Matrix collection\footnote{https://sparse.tamu.edu} as the $A$ matrix. The linear equation is solved by a sparse linear solver at the computing platform.

We ran the experiment 100 times and collected several performance metrics. In the next subsection, we present these metrics as well as the obtained results.

\subsection{Results and Discussion}

\begin{figure}[!htb]
\centering
\includegraphics[width=0.60\textwidth]{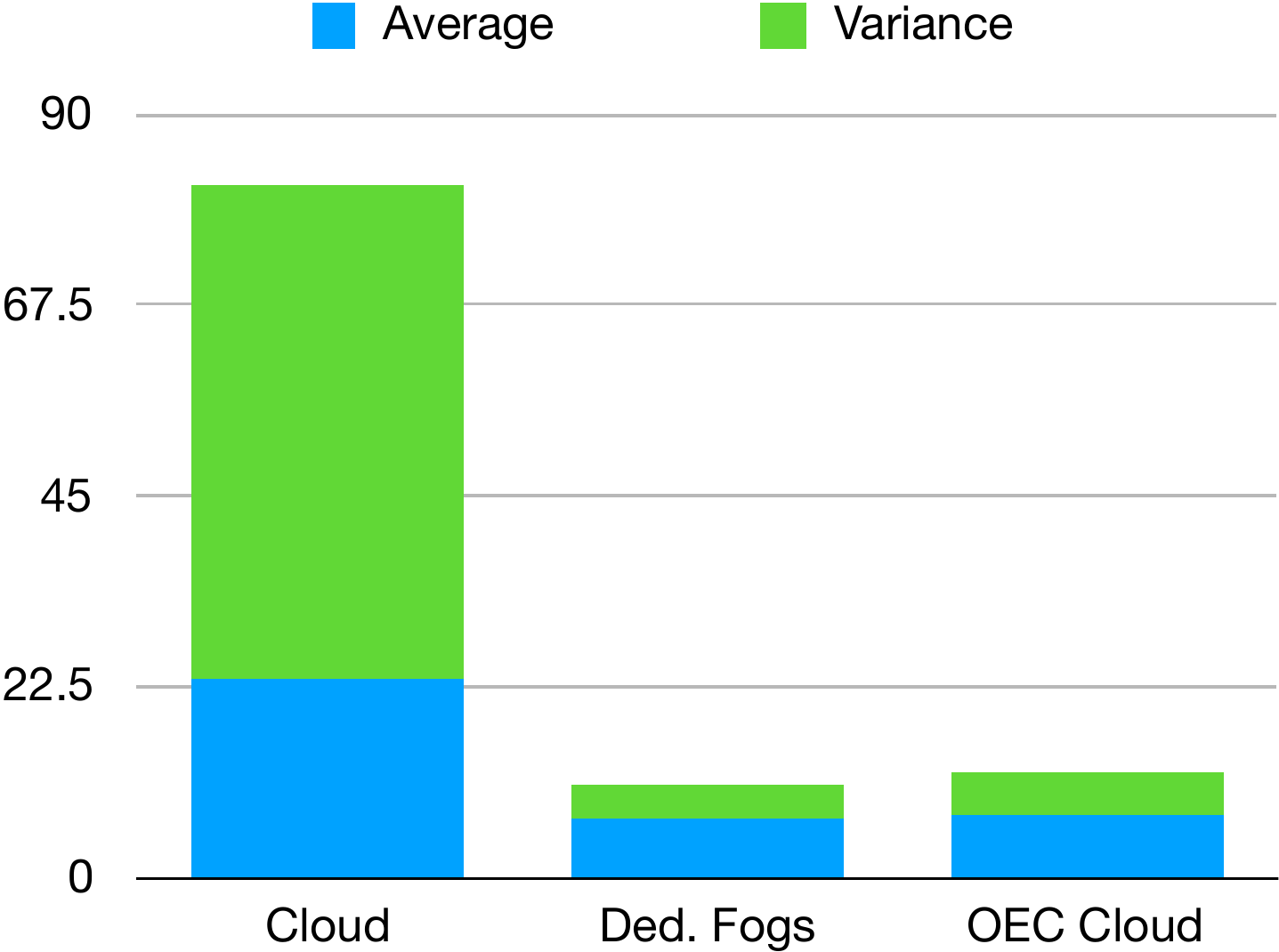}
\caption{Comparison of the averages and variances of the data upload times for cloud, dedicated fogs, and OEC Cloud}
\label{graph:uploadtime}
\end{figure}

The first metric used to compare the three computing platforms is the upload time, which is the time spent by the devices to upload the matrix A at the beginning of the experiment. Figure~\ref{graph:uploadtime} shows the average and variance of the data upload times  over several runs for the three studied scenarios. The dedicated fogs and OEC Cloud both perform very well with regard to the upload times because the resources are located at the edge in those cases, whereas the cloud requires the traversal of higher latency network connections and consequently high end-to-end application level latencies.

\begin{figure}[!b]
\centering
\includegraphics[width=0.60\textwidth]{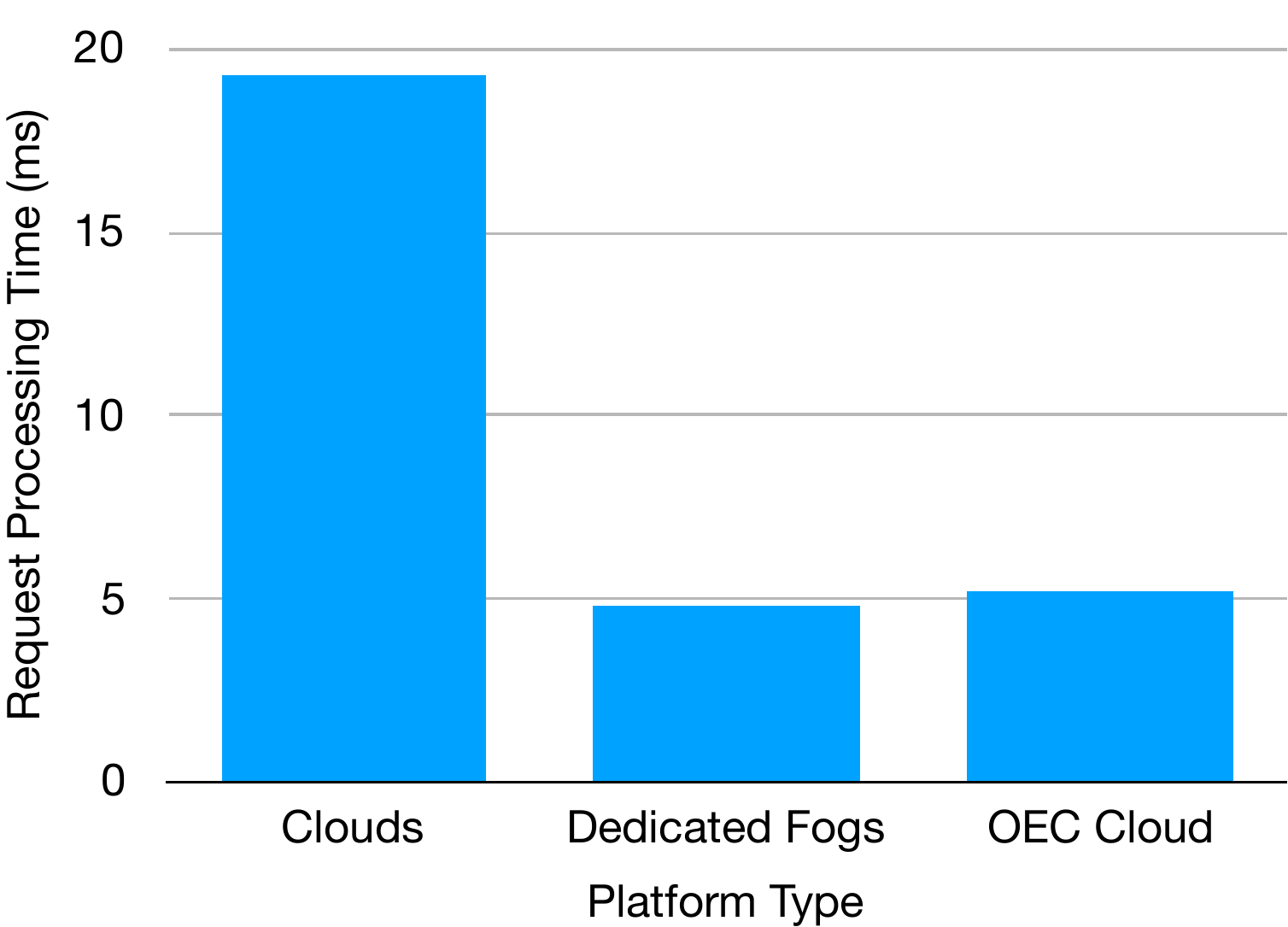}
\caption{Average request processing times for cloud, dedicated fogs, and OEC Cloud}
\label{graph:processtime}
\end{figure}

Once the matrix A is uploaded, end-user devices start sending the requests. We then compare the three computing platforms in terms of request processing time. The request processing time is the time taken for the end-user device to send the request (containing the vector $b$) and receive a reply from the computing platform. Figure~\ref{graph:processtime} shows the average request processing times measured from the end-user device side for the three computing platforms. It is clear from the figure that dedicated fogs and the OEC Cloud, which are both at the edge, outperform the cloud in terms of request processing time. We can also notice that the dedicated fogs performs slightly better (i.e., smaller request processing time).




Indeed, the disconnection and reconnection of the opportunistic resources is impacting the computation time in different ways. 
For instance, in our experiments, when a node from the OEC cloud is disconnected, it forgets the cached value of $A$ inverse so that it needs to recompute it at reconnection. 
\begin{figure}[!ht]
\centering
\includegraphics[width=0.60\textwidth]{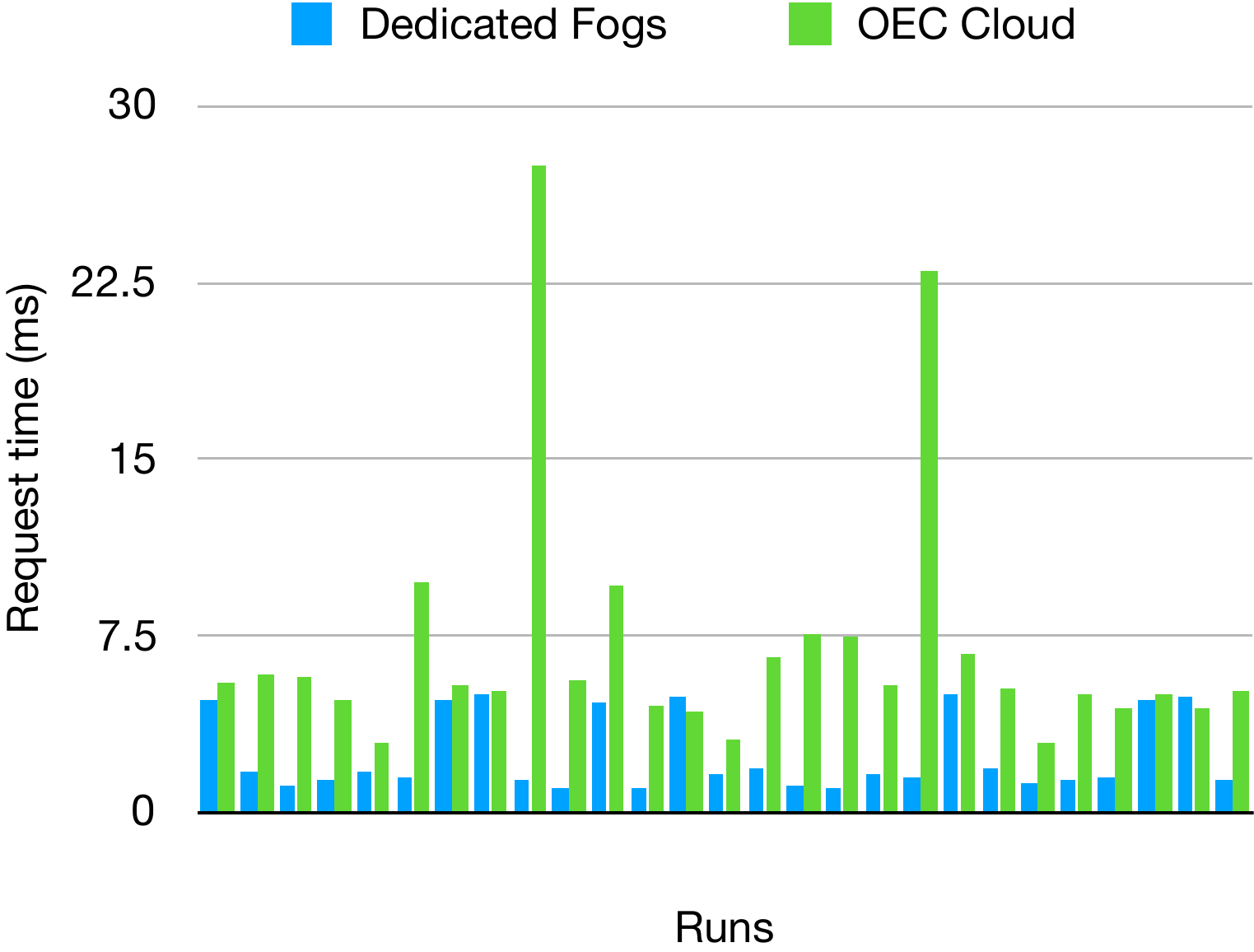}
\caption{Impact of disconnections on OEC Cloud in terms of request processing time}
\label{graph:dedicatedvsopp}
\end{figure}

Figure~\ref{graph:dedicatedvsopp} shows the impact of disconnection/reconnection on the request processing times for different runs. It clearly show that OEC clouds succeeds to lower the latency and keep it close to that of dedicated fogs where there are no disconnections. However, the request time is most of the time much lower than that of the cloud (less than~20ms as shown in Figure~\ref{graph:processtime}).

It is clear from these experiments that OEC makes the best trade-off between dedicated fogs and the cloud. On one hand, compared to Dedicated Fogs, OEC cloud could provide processing time and upload time that are close to that of dedicated fogs but with lower costs as there are deployment and maintenance costs. On the other hand, the OEC cloud definitely leverages the proximity to end-users to offer much lower latency and upload time compared to the regular cloud.


\section{Key Research Challenges in OEC Environments}
\label{challenges}

Opportunistic edge computing paradigm can be viewed as an extension of cloud computing into the edge. Hence, as expected, it inherits many of the research challenges of cloud and edge computing initiatives \cite{BARISurvey2013}. However, the self-scalability and the dynamic nature of OEC bring to the fore new challenges and research opportunities. Below we focus on the ones that are unique to OEC or have specific importance within the context of OEC.


\subsection{OEC Management Architectures} \label{sec:Architectures}
The opportunistic edge computing framework proposed in this paper delegates to the broker the hassle of building and managing the opportunistic cloud. This feature does not only eliminate the need for participants to directly negotiate among them to share resources but also further simplifies pricing and incentive management as all participants and service providers have to deal with a single entity, i.e., the broker.
The broker is hence responsible for all management-related tasks that are performed by the OEC framework as presented in Section~\ref{OECFramework}. However, how such framework should be deployed and implemented is still an open question. For instance, the framework elements can be implemented in a totally-centralized manner i.e., all decisions are taken at the same location like the cloud. 
It can be also deployed in a semi-centralized manner where a main management module cooperates with multiple local management systems. For instance, a local management system might be a system dynamically created at one of the edge devices to locally build and manage an opportunistic infrastructure and regularly report monitoring data and local decisions to the main management system.

In this context, many challenges could arise. For instance, a selection or an election process should be devised to select the device that should host the local management system based on several parameters like its location, performance and connectivity. Synchronization schemes could also be proposed to ensure that the main and local management systems have the same view of the infrastructure and make consistent decisions. 
The performance of such management architectures is still to be evaluated and eventually compared to~the~totally centralized scheme.



\subsection{Resource Management in OEC-based Environments} \label{sec:ResourceManagement}
Resource management is a central task in IT environments that refers to the efficient and effective deployment and allocation of the infrastructure's resources.
With OEC, resource management becomes more challenging as it needs to be performed not only with regular relatively stable resources (e.g., data centers, microclouds) but also with resources lying in devices that are heterogeneous  (i.e., with different processing and networking capacities), moving over time (i.e., the network topology is continuously changing over time), and available for only a limited time duration (i.e.,~the~available resource pool can vary with time). As a result, several new challenges arise with the OEC model.
\begin{itemize}
		 \item {\em OEC construction and monitoring}: Building opportunistically an edge cloud using the available devices requires designing new resource discovery and construction schemes able to identify the available resources (e.g.,~devices) and carefully select the ones to be included in the opportunistic edge infrastructure. Hence, new algorithms should be developed to dynamically select devices based on relevant metrics such as their computing capacity, predicted lifetime, reliability, connectivity and bandwidth performance. Furthermore, OEC monitoring techniques must be proposed in order to detect in a timely manner resource usage, devices' performance and availability, and the potential changes in the underlying infrastructure (e.g.,~new devices, disconnecting devices, failures). It is clear that monitoring such opportunistic infrastructures calls for novel data collection schemes that are able to minimize monitoring overhead (e.g.,~monitoring traffic) and, at the same time, to reduce the statistics' collection time.
		
		
		\item {\em Resource allocation in OEC}: Resource allocation has been extensively studied in the context of data centers \cite{ZhaniIM2013,VL2,Oktopus} and distributed infrastructures \cite{AmokraneTCC13,ZhangJSAC13,ADNAN2012}. However, these infrastructures have relatively unchanging topologies over time compared to OEC. With OEC, resources are allocated over physical topologies that are highly variable over time and space, especially at the edges. This makes resource allocation more challenging as it should take into consideration more constraints like the node lifetime, direction, speed, energy and reliability in addition to the usual parameters like the node computing and networking capacities. In this context, new allocations schemes should be proposed to take into account these constraints and continuously ensure that the resources promised to service providers is always available. For instance, when a participant intends to quit the opportunistic edge infrastructure, resource migration strategies could be proposed to make sure that services (i.e., usually hosted in virtual machines or containers) can be dynamically migrated to other active participants.
		
		  \item {\em Network support for edge provisioning}: In OEC, resources are obtained from the constructed opportunistic infrastructure, which encompasses the neighboring devices (e.g., smart phones, PCs and cloudlets) and the network connecting them. Therefore, routes used by traffic flows within this opportunistic network should be optimized according to the variable locations of the edge participants and their bandwidth capacities. One way worth exploring in order to solve this problem is to leverage technologies like software defined networking to optimally and dynamically route the traffic based on the current characteristics of the devices (e.g.,~performance, utilization, bandwidth) and the provisioning goals. 

\end{itemize}
\subsection{Fault Tolerance and Reliability} \label{sec:FaultToleranceSurvivability}
Fault-tolerance and reliability are key design goals that have a direct impact on the overall system performance and end-users' satisfaction. OEC has the ability to perform edge resource provisioning using resources that are contributed by participants. Therefore, resources offered by OEC can fail if any of them break the obligation to OEC. 

One approach is to use a soft state paradigm just like what is proposed in the context of fogs and cloudlets \cite{satyanarayanan2009case}. Soft state stipulates that cloudlets or less reliable devices should only contain caches or copies of codes. As such, the failure or the disconnection of the device will not incur any data or code loss. Furthermore, VM and data replication techniques could also be used to ensure the continuity of the service in case of failures. In this context, a major challenge would be to develop efficient fault-tolerance management techniques that dynamically determine the number of replicas and caches to create and then place them in the cloud and in the OEC edge cloud. 
Unlike existing work (e.g., \cite{zhani2015survivability,RabbaniIEICE13,ZhangVenice2014}), these techniques should not only satisfy the availability and performance requirements stipulated by service providers but also consider the distributed nature of the OEC infrastructures (i.e., remote clouds, edge cloud with micro-clouds and participants' devices), the mobility of the devices and their different availability durations. The ultimate goal would be to ensure high resource availability and to mitigate failure impact on the deployed applications.
		

\subsection{Pricing and Incentives} \label{sec:PricingIncentives}
In the context of OEC, pricing is a fundamental challenge for brokers and service providers as it will affect their costs, revenues and profits. 
Brokers pool resources from participants and then offer it to service providers. In other words, the broker is ``buying" resources from participants and then sell it to service providers. As a result, the broker needs to provide incentives to participants to urge them to contribute with their resources. 
\begin{itemize}
		\item {\em Incentives}: An OEC-based system would incentivize the participants to lend their resources by advertising the ``rewards" offered by the system for certain types of resources at different network localities. These rewards could be monetary or simply services to be offered to the participant in exchange for their resources. 
		For instance, recent work ~\cite{goyal2011collective} proposed to pay participants in a virtual currency when they contribute with their resources and they can use the same currency to use some services.
		The rewards could also be derived based on the historical and the forecasted demands for the specific types of resources and their availabilities in the network vicinity. For instance, a resource type that is available in plenty would have very low price and would not attract additional contributions, whereas scarce resources would attract contributions because their price would be high.  More incentive mechanisms are hence to be designed and developed by brokers in order to further attract more participants to the OEC and guarantee that resources are available whenever and wherever they are needed. 
		Furthermore, validation mechanisms are necessary to track the achieved performance and resource usage during deployment and reevaluate the incentives offered to the resource provider.
		\item {\em Pricing}: Pricing models should define the prices for the resources offered by the broker to the service providers. While the ``pay-as-you-go" pricing model widely used in cloud computing environments could be interesting for OEC, it might not be sufficient as the resources at the edges may not be permanent and can be scarce and highly requested at some periods of time. Hence, a compelling challenge would be to devise novel pricing models that dynamically estimate prices for the resources discovered at the edge. Such a model should take into consideration several parameters that are unique to OEC environments like the type of the discovered resources available at the edges, their potential variable costs as well as their different reliability and availability over~time. 
		
Auction-based spot pricing could also be an appealing option for OEC. 
Such pricing is already adopted by Amazon EC2 Spot instances \cite{AmazonEC2SpotInstances} where users are allowed to bid on the spare Amazon EC2 computing capacity. Amazon decides the spot price and only accepts bids above it. Similarly, in OEC, service providers could be allowed to bid on the resources available at the edge infrastructure. In this case, an interesting challenge would be to dynamically adjust the minimal spot price based on the availability and scarcity of the resources and also on the demand in order to maximize the broker's expected profits or to ensure fairness between service providers.

		
		%
		
		
\end{itemize}

\subsection{Security and Trust Management} \label{sec:SecurityTrustManagement}

Security has to be always at the heart of any IT environment. 
OEC involves several stakeholders that are sharing resources among them.  As a result, security has to be taken into account at different levels from user management to resource sharing, to access management, to the security of applications and services that are deployed and offered to end users. In OEC, resources are sliced by the broker and offered to multiple service providers, virtualization and container technologies can be a valuable tool to provide  different isolated environments to service providers. In this context, developing lightweight virtualization architectures capable of ensuring a high level of performance and security isolation between the resource slices is an important problem as shown in some recent studies \cite{Felter2015}. Furthermore, OEC resources might be owned by participants (who might be end-users). This makes it different from the traditional cloud computing model where the infrastructure is owned or managed by the same entity (e.g.,~Amazon~EC2, Google Cloud Platform) \cite{BARISurvey2013}. As a result, security issues become more challenging as the broker may not have full control over all the managed resources and the technologies deployed on participants' devices. 


One particular challenge that is still also to be explored is trust management between the broker and the participants.
Trust has been defined in \cite{Grandison2000} as \textsl{the  firm  belief  in  the  competence  of  an  entity  to  act dependably, securely, and reliably within a specified context}. In OEC, novel trust management schemes should be developed to quantify broker-to-participant trust and evaluate how far the broker can rely on participants and use the resources they are offering. Such schemes should not be limited to using credentials but also extended to analyze previous experiences with the participants and take more informed decisions taking into account expectations in terms of performance, availability and security.

\section{Concluding Remarks}

In this paper, we present an Opportunistic Edge Computing (OEC) framework for creating and managing
resource pools built using participant contributions and available resources at the access networks (e.g.,
Cloudlets). One of the important benefits of OEC is its self-scaling nature, where the end users  are incentivized
to actively contribute to scale up the edge computing resource pool. The OEC framework is
responsible for
building, managing, and slicing this pool of resources while meeting the performance and reliability
objectives.
To realize the vision of OEC, we have
identified many important research challenges ranging from the design of the overall OEC management architecture to the development
of pricing and incentive schemes.



\end{document}